%% file: LLRmdrag.TEX
\documentstyle[12pt,iopconf]{article}
\input{prepicte}
\input{pictex}
\input{postpict}

\begin{document}
\newcommand{\be}{\begin{equation}}
\newcommand{\ee}{\end{equation}}
\newcommand{\bea}{\begin{eqnarray}}
\newcommand{\eea}{\end{eqnarray}}

\title{Lunar Laser Ranging - A Comprehensive Probe \\ of Post-Newtonian Gravity}
\author{Kenneth Nordtvedt\\{\it Northwest Analysis, 118 Sourdough Ridge, Bozeman MT 59715 USA}\\{\it kennordtvedt@one800.net}}

\begin{center}
\section*{Abstract}
\end{center}
Using more than 30 years of lunar laser ranging (LLR) data, the lunar orbit has been modeled and fit with several millimeters precision.  This fit provides comprehensive confirmation of the general relativistic equations of motion for solar system dynamics, achieving several key tests of Einstein's tensor theory of gravity, and strongly constraining presence of any additional long range interactions between bodies. Earth and Moon are found to fall toward the Sun at rates equal to better than a couple parts in $10^{13}$, confirming both the universal coupling of gravity to matter's stress-energy tensor, and gravity's specific non-linear coupling to itself. The expected {\it deSitter} precession (with respect to the distant 'fixed' stars) of the local inertial frame moving with the Earth-Moon system through the Sun's gravity is confirmed to 3.5 parts in $10^3$ precision ($\pm\;.07\;mas/year$), and Newton's gravitational ``constant'' indeed shows no cosmological time variation at the part in $10^{12}\;per\;year$ level. Most all of the $1/c^2$ order, post-Newtonian terms in the N-body equations of motion --- motional, gravito-magnetic, non-linear, inductive, etc.  --- contribute to the measured details of the lunar orbit, so LLR achieves near-completeness as a gravity experiment and probe. The precision of these measurements, especially those connected with lunar orbit frequencies and rates of change of frequencies, will further improve as LLR observations continue into the future with use of latest technologies.

\clearpage  

\section{Introduction.}

The precise fit of the lunar laser ranging (LLR) data to theory yields a number of the most exacting tests of Einstein's field theory of gravity, General Relativity, because almost any alternative theory of gravity predicts a number of changes (from that produced by General Relativity) in the lunar orbit which would be readily detected in the LLR data. Some of the most interesting and fundamental of such theory-dependent effects, and which are particularly well-measured by LLR, include 1) a difference in the free fall rate of Earth and Moon toward the Sun due to gravity theory's non-linear structure acting on the gravitational binding energy within the Earth, 2) a time variation of Newton's gravitational coupling parameter, $G\rightarrow G(t)$, related to the expansion rate of the universe, and 3) precession of the local inertial frame (relative to distant inertial frames) because of the Earth-Moon system's motion through the Sun's gravity.

Measurements of the round trip travel times of laser pulses between Earth stations and sites on the lunar surface have been made on a frequent basis ever since the Apollo 11 astronauts placed the first passive laser reflector on the Moon in 1969.  Today about 15,000 such range measurements are archived and available for use by analysis groups wishing to fit the data to theoretical models for the general relativistic gravitational dynamics of the relevant bodies, the speed of light function in the solar system, tidal distortions of Earth and Moon, atmospheric corrections to light propagation, etc.  An individual range measurement today has precision of about a centimeter (one-way), but a new generation observing program plans to improve this range measurement precision down to a millimeter.  Because of the large number of range measurements, some of the key length parameters which describe the lunar orbit are already estimated with precision of a few millimeters, and key lunar motion frequencies to fractional precisions of a few parts in $10^{12}$.   

Because both the Earth's mass and that of the Moon are sufficiently large, the orbits of these bodies can be modeled as single orbital "arcs" extending over three decades through time. The complete model used to fit the many range measurements contains in excess of a hundred parameters $P_m$ which are optimally adjusted from their nominal model values $P_m^{(o)}$ by amounts $\delta P_m=P_m-P_m^{(o)}$ determined in a weighted least squares fit type procedure
\[
Minimize\hspace{.5in}\sum_{i,j=1}^N\,W_{ij}\,\sum_{m,n=1}^M \left[f(m)_i\,\delta P_m\,-\,r_i\right]\left[f(n)_j\,\delta P_n\,-\,r_j\right]
\]
with the $N$ range measurements being identified by the labels $i$ and $j$, and the $M$ model parameters being identified by the labels $m$ and $n$. $W_{ij}$ are the weightings given to each measurement (pair) and are usually taken to be diagonal in $ij$ and inversely proportional to the square of inferred measurement errors; the {\it residuals} $r_i$ are the differences between observed and calculated range values, $r_i=R_{obs}(t_i)-R_{calc}(t_i)$; and the remaining functions $f(m)_i$ are the parameter {\it partials} which give the sensitivity of the modeled (calculated) range to change in each model parameter value
\[
f(m)_i\;=\;\frac{\partial R_{calc}(t_i)}{\partial P_m}
\]
evaluated at the time $t_i$ of the $ith$ range measurement.

Among the very many model parameters, the information needed for testing relativistic gravity theory is concentrated in only a handful of orbital features.  The needed orbital parameters are connected with four key oscillatory contributions to the lunar motion, the {\it eccentric, evective, and variational} motions and the {\it parallactic inequality}, which are illustrated in Figure 1. The eccentric motion produces an oscillatory range contribution proportional to $cos(A)$, $A$ being anomalistic (eccentric) phase, and is a natural and undriven perturbation of circular motion.  The {\it variation} is driven by the Sun's leading order quadrupolar tidal field and produces a range contribution proportional to $cos(2D)$, $D$ being synodic phase from {\it new moon}.  The {\it parallactic inequality} is driven by the Sun's next order octupolar tidal field, and its range perturbation $cos(D)$ has monthly period. The {\it evection} is a hybrid range perturbation proportional to the eccentric motion as modified by the {\it variation} and having time dependence $cos(2D-A)$.  The eccentric and evective motions, which alter the times of eclipses, were discovered by the anchients;  the variation and parallactic inequalities, which do not alter times of eclipses, were only found during and after the era of Newton.

The amplitude  of the parallactic inequality, $L_{PI}$, is unusually sensitive to any difference in the Sun's acceleration rate of Earth and Moon \cite{norc}.  The frequency of the eccentric motion, the {\it anomalistic} frequency $\dot{A}$, when compared to other lunar frequencies determines the precession rate of the Moon's perigee.  This rapid precession, which completely rotates the orbit's major axis in about 8.9 years, is primarily driven by the Sun's tidal acceleration, but there is a leading order relativistic contribution to this precession rate interpreted as an actual rotation of the local inertial frame, the {\it deSitter precession}, resulting from motion through the Sun's field of gravity. From measurement of the time rates of change of the Moon's {\it anomalistic} and {\it synodic} frequencies, $\ddot{A}$ and $\ddot{D}$,  a rather clean measurement can be made of a time rate of change of Newton's coupling parameter $G$.  The Earth-Moon range model can be expressed in terms of these primary contributions
\[
r(t)\;=\;L_o\;-\;L_{ecc}\,cos(A)\;-\;L_{evc}\,cos(2D-A)\;-\;L_{var}\,cos(2D)\;-\;L_{PI}\,cos(D)\;+\;...
\]
with phases advancing as $A=A_o+\dot{A}(t-t_o)+\ddot{A}(t-t_o)^2/2+...$ and similarly for synodic phase $D$.  The LLR measurement of $L_{PI},\;\dot{A},\;\dot{D},\;\ddot{A},$ and $\ddot{D}$ are the foundations for the gravity theory tests.    

\begin{figure}
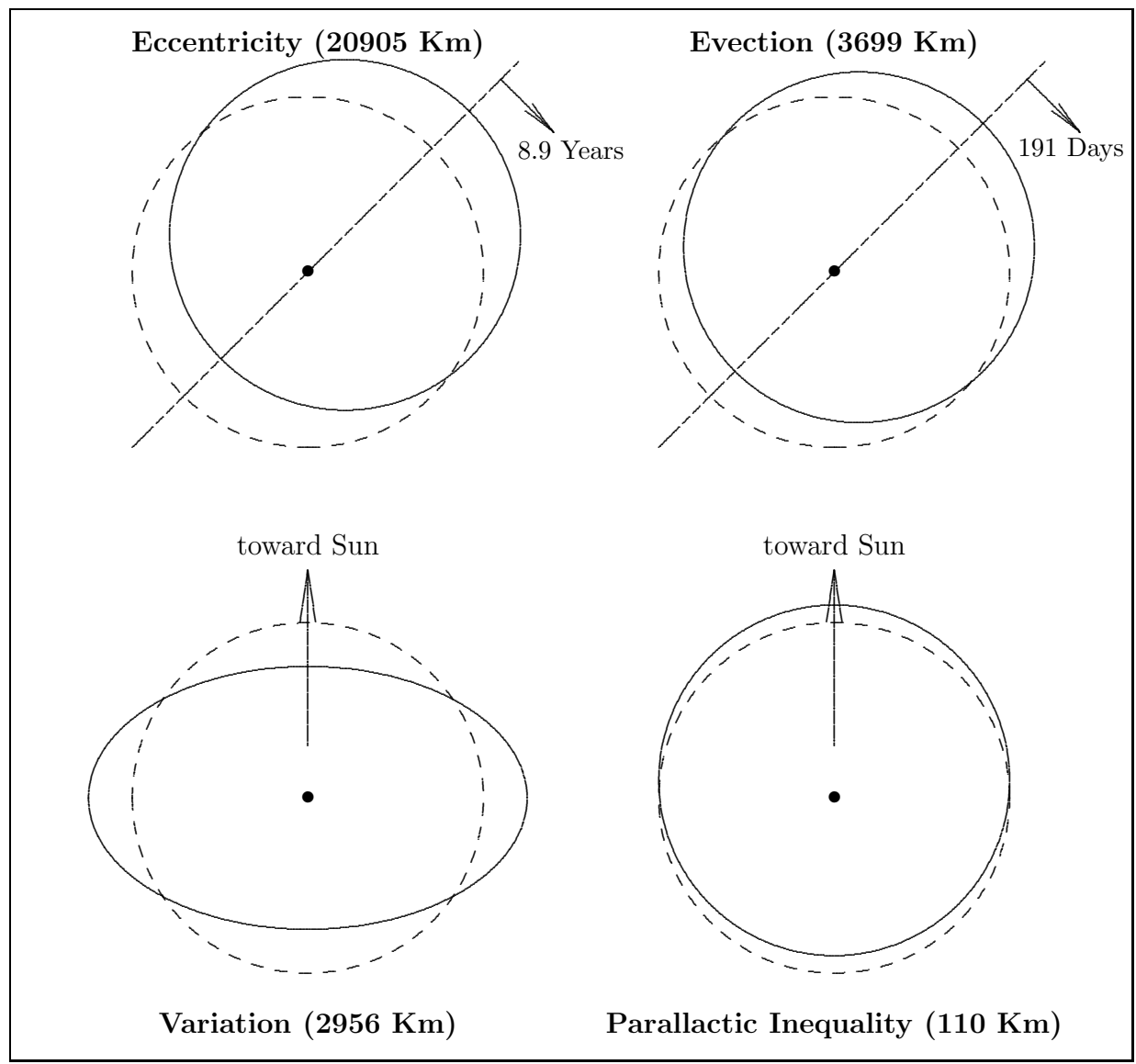

\beginpicture
\setcoordinatesystem units <1in,1in>
\setplotarea x from -3.2 to 3.2, y from 0 to 6
\putrectangle corners at -3.2 0 and 3.2 6
\plotheading {\Large \bf Lunar Orbit's Four Main Perturbations}
\put {$\bullet$} at 1.5 1.5
\put {$\bullet$} at -1.5 1.5
\put {$\bullet$} at -1.5 4.5
\put {$\bullet$} at 1.5 4.5
\ellipticalarc axes ratio 1.25:.75 360 degrees  from -2.75 1.5 center at -1.5 1.5
\circulararc 360 degrees from 1.5 2.6 center at 1.5 1.6
\startrotation by .707 .707 about 1.5 4.5
\circulararc 360 degrees from .7 4.5 center at 1.7 4.5
\stoprotation
\startrotation by .707 .707 about -1.5 4.5
\circulararc 360 degrees from -.2 4.5 center at -1.2 4.5
\stoprotation
\arrow <.3in> [.15,.3] from -1.5 1.8 to -1.5 2.8
\arrow <.3in> [.15,.3] from 1.5 1.8 to 1.5 2.8
\arrow <.2in> [.12,.3] from 2.6 5.6 to 2.9 5.3
\put {\small 191 Days} at 2.85 5.2
\arrow <.2in> [.12,.3] from -.4 5.6 to -.1 5.3
\put {\small 8.9 Years} at 0 5.2
\put {toward Sun} at -1.5 2.95
\put {toward Sun} at 1.5 2.95
\put {\it \bf Variation (2956 Km)} at -1.5 .2
\put {\it \bf Parallactic Inequality (110 Km)} at 1.5 .2
\put {\it \bf Eccentricity (20905 Km)} at -1.5 5.8
\put {\it \bf Evection (3699 Km)} at 1.5 5.8
\setdashpattern <12pt, 2pt, 4pt, 2pt, 4pt, 2pt>
\setlinear \plot .5 3.5 2.7 5.7 /
\setlinear \plot -2.5 3.5 -.3 5.7 /
\setdashes
\circulararc 360 degrees from -2.5 1.5 center at -1.5 1.5
\circulararc 360 degrees from 2.5 1.5 center at 1.5 1.5
\circulararc 360 degrees from 1.5 3.5 center at 1.5 4.5
\circulararc 360 degrees from -1.5 3.5 center at -1.5 4.5
\endpicture
\caption{Four lunar orbit perturbations from a nominal circular orbit (dotted) are shown. They produce oscillatory Earth-Moon range terms: the eccentric oscillation $\sim cos(A)$, the variation oscillation $\sim cos(2D)$, the parallactic inequality oscillation $\sim cos(D)$, and the evective oscillation $\sim cos(2D-A)$, with respective amplitudes indicated. Key tests of general relativity are achieved from precise measurements of amplitudes or phase rates of these perturbations. Measurement of the amplitude of the parallactic inequality determines whether Earth and Moon fall toward the Sun at same rate. Measurements of the synodic phase $D$ and anomalistic (eccentric) phase $A$ rates and rate of change of these rates determine the deSitter precession of the lunar orbit and time rate of change of Newton's $G$.} 
\end{figure}

\section{Dynamical Equations For Bodies, Light, and Clocks.}  

LLR comprehensively tests the $1/c^2$ order, gravitational N-body equations of motion which analysis groups integrate to produce orbits for Earth, Moon, and other relevant solar system bodies.  The Sun-Earth-Moon system dynamics is symbolically illustrated in Figure 2, with the rest of the solar system bodies sufficiently considered at the Newtonian level of detail. The Earth moves with velocity $\vec{V}$ and acceleration $\vec{A}$ with respect to the Sun, while the Moon is moving at velocity $\vec{V}+\vec{u}$ and acceleration $\vec{A}+\vec{a}$.  (If {\it preferred frame} effects were to be considered for cases when gravity is not locally Lorentz-invariant, the Sun's cosmic velocity $\vec{W}$ also becomes involved.) \cite{nor73}  There are a variety of post-Newtonian forces acting on Earth and Moon by the Sun, by each other, and on themselves (self forces) which are dependent on these general motions.  Included in these are {\it non-linear} gravitational forces for which each mass element of the Earth and Moon experiences forces due to the interactive effect of the Sun's gravity with the other mass elements of the same body, or of the other neighboring body.  The accelerations of individual mass elements of Earth also induce accelerations on the other mass elements of Earth, and similarly with the Moon.  Acceleration of Earth induces an acceleration of the Moon.  Altogether, these $1/c^2$ order accelerations produce a rich assortment of modifications of the Earth-Moon range which LLR can measure.

\begin{figure}
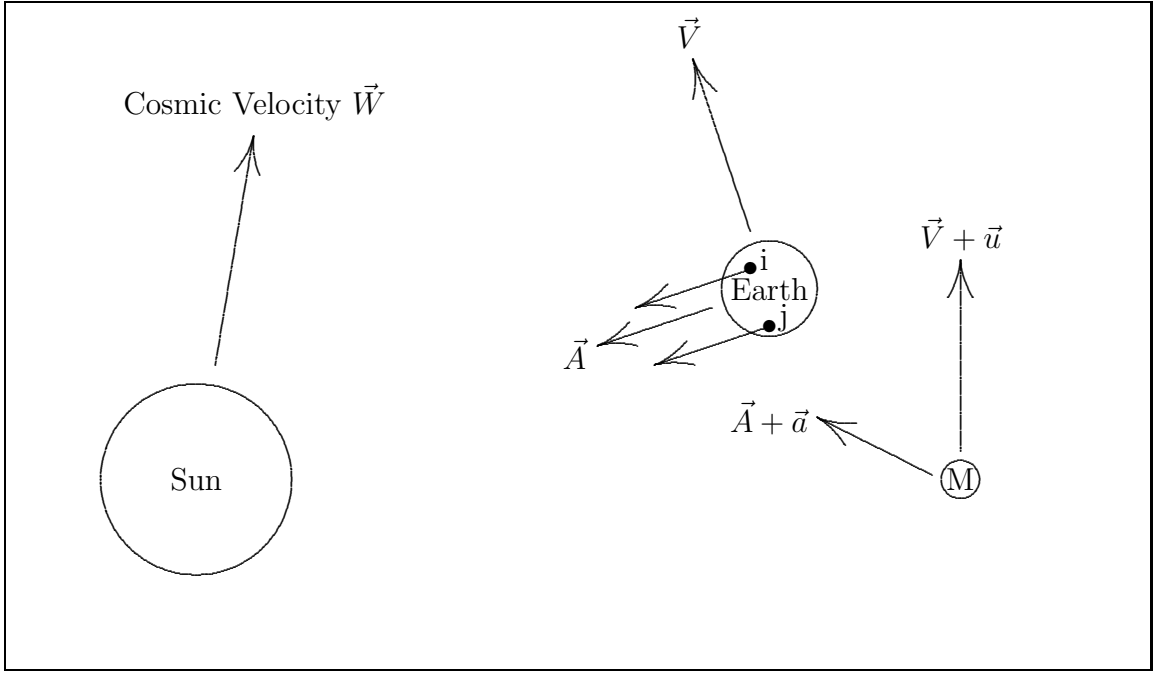

\beginpicture
\setcoordinatesystem units <1in,1in> point at 0 0
\setplotarea x from -3.2 to 3.2, y from 0 to 3.5
\put {Sun} at -2 1
\put {Earth} at 1 2
\put {M} at 2 1
\circulararc 360 degrees from 1.25 2 center at 1 2
\arrow <.2in> [.2,.67] from .9 2.3 to .6 3.2
\put {$\vec{V}$} [b] at .58 3.26
\arrow <.2in> [.2,.67] from .7 1.9 to .1 1.7
\arrow <.2in> [.2,.67] from .9 2.1 to .3 1.9
\arrow <.2in> [.2,.67] from 1 1.8 to .4 1.6
\put {i} [l] at .95 2.15
\put {j} [l] at 1.05 1.85
\put {$\bullet$} at .9 2.1
\put {$\bullet$} at 1 1.8
\put {$\vec{A}$} [r] at .04 1.66
\circulararc 360 degrees from -2.5 1 center at -2 1
\arrow <.2in> [.2,.67] from -1.9 1.6 to -1.7 2.8
\put {Cosmic Velocity $\vec{W}$} [b] at -1.695 2.88
\circulararc 360 degrees from 2.1 1 center at 2 1
\arrow <.2in> [.2,.67] from 2 1.15 to 2 2.15
\arrow <.2in> [.2,.67] from 1.85 1.025 to 1.25 1.325
\put {$\vec{V}+\vec{u}$} [b] at 2 2.2
\put {$\vec{A}+\vec{a}$} [r] at 1.20 1.325
\putrectangle corners at -3 0 and 3 3.5 
\plotheading {\Large Velocities and Accelerations of Sun, Earth, and Moon}
\endpicture
\caption{\small When formulating the Earth-Moon dynamics in the solar system barycentric frame, there are post-Newtonian force terms acting between Sun, Earth and Moon which depend on either the velocity or acceleration vectors of both the Earth and Moon.  Body {\it self-accelerations} also result from the inductive inertial forces acting between the mutually accelerating mass elements ($i,\;j$) within each of these bodies. The intrinsic non-linearity of gravity also produces net external forces on these bodies proportional to not only the presence of other bodies, but also to their internal gravitational binding energies.  The motional, accelerative, and non-linear contributions to the three body system's dynamics, taken collectively, make LLR a comprehensive probe of the post-Newtonian dynamics of metric gravity in the general case.  If the dynamics is not locally Lorentz invariant, then the velocity $\vec{W}$ of the solar system through the cosmos leads to novel forces  and resulting observable effects in LLR proportional to $\vec{W}$ (or its square); but such effects have not been seen.}   
\end{figure}

The N-body equation of motion in metric gravity has been formulated in the literature for the completely general case \cite{wnor73}.  Not observing  any violations of local Lorentz-invariance or breakdown of conservation laws in solar system gravity, I here specialize consideration to the fully conservative, locally Lorentz-invariant, lagrangian-based gravitational equation of motion (plus cosmological variation of Newton's $G$). For N bodies in general motion and configuration, and valid for a broad class of plausible metric theories of gravity, scalar-tensor theories in particular, the order $1/c^2$ equations of motion for these $N$ bodies take the form           

\bea
&A&\hspace{.4in}\vec{a}_i\;=\;\left(1+\frac{\dot{G}}{G}(t-t_o)\right)\left(\frac{M(G)}{M(I)}\right)_i\;\vec{g}_i \nonumber \\
&B&\hspace{.4in}-\;\beta^*\;\sum_{j\neq i}\left(\sum_{k\neq i}\frac{\mu_k}{r_{ik}}
+\sum_{k\neq j}\frac{\mu_k}{r_{jk}}\right)\;\vec{g}_{ij} \nonumber \\
&C&\hspace{.4in}+\;(2\gamma+2)\sum_{j\neq i}\vec{v}_i\times(\vec{v}_j\times\vec{g}_{ij})
 \nonumber \\
&D&\hspace{.4in}+\;\frac{1}{2}\sum_{j\neq i}\left[(2\gamma+1)v_i^2\:+\:(2\gamma+2)v_j^2\:-\:3(\vec{v}_j\cdot\hat{r}_{ij})^2\right)\:\vec{g}_{ij}\:-\:(4\gamma+2)\left(\vec{g}_{ij}\cdot\vec{v}_j(\vec{v}_j-\vec{v}_i)\:+\:\vec{g}_{ij}\cdot\vec{v}_i\vec{v}_i\right] \nonumber \\
&E&\hspace{.4in}+\;\frac{1}{2}\sum_{j\neq i}\frac{\mu_j}{r_{ij}}\left[(4\gamma+3)\vec{a}_j\:+\:\vec{a}_j\cdot\hat{r}_{ij}\hat{r}_{ij}\right] \nonumber \\
&F&\hspace{.4in}-\;\frac{1}{2}v_i^2\vec{a}_i\:-\:\vec{a}_i\cdot\vec{v}_i\vec{v}_i\:-\:(2\gamma+1)\sum_{j\neq i}\frac{\mu_j}{r_{ij}}\vec{a}_i
\eea
with $\vec{v}_i\:=\:d\vec{r}_i/dt$, $\vec{a}_i\:=\:d\vec{v}_i/dt$,  $r_{ij}\:=\:|\vec{r}_i-\vec{r}_j|$, and $i,j,k\:=\:1\;...\:N$.  The speed of light factor $1/c^2$ has been set equal to 1 in lines $B\;...\;F$ to simplify presentation. The body gravitational mass strengths $\mu_i\:=\:GM(G)_i$ are indicated along with the Newtonian acceleration vectors 
\[
\vec{g}_{ij}\;=\;\frac{\mu_j}{r_{ij}^3}\vec{r}_{ji}\hspace{.3in}and\hspace{.3in}\vec{g}_i\;=\;\sum_{j\neq i}\vec{g}_{ij} 
\]
$\gamma$ and $\beta$ (with $\beta^*=2\beta-1$) are two {\it Eddington} parameters which quantify deviations of metric gravity theory from Einstein's pure tensor theory in which both these parameters equal one.  The several lines of this total equation of motion warrant individual descriptions and brief discussions.

Line $A$.  If the metric theory {\it Eddington} parameters $\gamma$ and $\beta$ differ from their general relativistic values $\gamma_{gr}=\beta_{gr}=1$, application of the equation of motion relativistic corrections from lines $B$ through $F$ to a body's internal gravity finds that the gravitational to inertial mass ratio of a celestial body depends on its gravitational self-energy content \cite{norb}
\be
\frac{M(G)}{M(I)}\;=\;1\:-\:(4\beta-3-\gamma)\;\frac{G}{2Mc^2}\int\frac{\rho(\vec{x})\rho(\vec{y})}{|\vec{x}-\vec{y}|}\:d^3x\:d^3y\;+\;order\:1/c^4
\ee
Another way to view the above ratio is in terms of a spatially varying gravitational coupling parameter $G$
\[
G(\vec{r},\,t)\;\cong\;G_{\infty}\,\left[1-(4\beta-3-\gamma)\,U(\vec{r},\,t)/c^2\,\right]
\]
in which a body with a significant part of its mass-energy coming from its gravitational binding energy experiences the additional acceleration 
\[
\delta\vec{a}_i\;\cong\;-\,\frac{\partial\,lnM_i}{\partial\,G}\,c^2\,\vec{\nabla}G
\]
with the leading gravitational energy contribution to body mass being Newtonian contribution
\[
\frac{\partial\,M}{\partial\,lnG}\;=\;-\frac{G}{2c^2}\int\frac{\rho(\vec{r})\rho(\vec{r}\,')}{|\vec{r}-\vec{r}\,'|}\,d^3r\,d^3r'
\]
When cosmological equations from a metric theory are considered, Newton's coupling parameter $G$ will also generally be found to vary in time in proportion to the Hubble expansion rate of the universe
\be
\frac{\dot{G}}{G}\sim\;(4\beta-3-\gamma)\;H
\ee
The presently most precise way to measure any deviation of $\beta$ from its general relativistic value is through measurement of the $M(G)/M(I)$ ratio of Earth using LLR data.

Line $B$. Gravity couples to itself, thereby producing  non-linear gravitational forces among and between bodies.

Line $C$. Just as pairs of moving charges generate magnetic forces between themselves in proportion to the velocities of both charges, pairs of moving masses generate {\it gravito-magnetic} forces between themselves.  This force acts between the mutually moving Earth and Moon and contributes to the necessary Lorentz-contraction of the lunar orbit as viewed from the solar system barycenter.

Line $D$. Masses in motion both produce and couple to gravitational fields differently than masses at rest. The package of velocity-dependent acceleration terms in this line plus line $C$ lead to local Lorentz-invariance of gravity. Any further modifications of this package (beyond the $\gamma$-dependence) will lead to additional terms in the equation of motion with one or two powers of body velocities being replaced by the velocity $\vec{W}$ of the solar system relative to the universe {\it preferred frame}.  A variety of preferred frame effects which would then result have been empirically ruled out in LLR and other solar system observations \cite{norw}.

Line $E$. Accelerating masses generate inductive gravitational forces on other proximite masses.  

Line $F$. The inertia of a mass is altered by its motion and by its proximity to other masses.  The combination of terms from this line plus line $E$ are necessary in order that a body's gravitational self-energy contributes to its total inertial mass in accord with special relativity's prescription $M=E/c^2$. This modification of inertia is part of the $M(G)/M(I)$ calculation for a celestial body.
    
LLR measures the round trip time of propagation of light between two separate body trajectories, and this measurement is made by a specific clock moving on a particular trajectory. So in the solar system barycentric and spatially isotropic coordinates employed to express the body equations of motion given by Equation (1), there are also requirements for the post-Newtonian modifications to the light coordinate speed function and to the clock rates, these respectively being
\be
c(\vec{r},\,t)\;\cong\;c_{\infty}\:\left[1\,-\,(1+\gamma)\;U(\vec{r},\,t)/c^2\,\right] 
\ee
and  
\be
d\tau\;\cong\;dt\;\left[1\,-\,v^2/2c^2\,-\,U(\vec{r},\,t)/c^2\,\right]
\ee
in which $U(\vec{r})$ is the total Newtonian gravity potential function due to solar system bodies 
\be
U(\vec{r},\,t)\;=\;\sum_j\int \frac{G\rho(\vec{r}\,'(t))_j}{|\vec{r}-\vec{r}\,'(t)|}\;d^3 r'
\ee
Because the Earth moves in the solar system barycentric frame, and it rotates at rate $\vec{\nu}$, there must be two corrections applied to an Earth surface location $\vec{a}$; first there is the Lorentz-contraction of the extended body
\[
\delta\vec{a}\;\cong\;-\,\vec{a}\cdot\vec{V}\,\vec{V}/2c^2
\]
and because of special relativity's non-absolute nature of time simultaneity there is a further displacement of the rotating Earth surface locations 
\[
\delta\vec{a}\;\cong\;\vec{V}\cdot\vec{a}\;(\vec{\nu}\times\vec{a})/c^2
\] 
These light and clock equations, and special relativistic body distortion effects play only  supportive (but necessary) roles in fitting LLR data; the main science emerges from the body equations of motion as given by Equation (1).

\section{LLR's Key Science-Related Range Signals.}

Assocated with each feature of gravitational theory which is tested by LLR, there are specific range signals in the LLR data whose measurements yield the information about theory.  Several of these signals are here described.

\subsection*{Violation of Universality of Free-fall.} 

Because of gravitational self-energies (internal gravitational binding energies) in celestial bodies, they will generally possess gravitational to inertial mass ratios which differ from each other as indicated in line $A$ of Equation (1) and given by Equation (2).  But there are other ways in which bodies may accelerate at different rates toward other bodies.  Within the paradigm that forces between objects are carried by a field, an additional long-range interaction in physical law generates a force between bodies $i$ and $j$ which will typically have the static limit form 
\be
\vec{f}_i\;=\;K_i\;\vec{\nabla}_i\:\frac{K_j}{r_{ij}}\:e^{-\mu\:r_{ij}}
\ee
The bodies' coupling  strengths $K_i$ and $K_j$, except in  special cases such as metric scalar-tensor gravity in which $K_i\sim M_i$, will be attributes of the bodies which are different than total mass-energy ({\it non-metric} coupling); and the dependence on distance of this force will be either inverse square if the field is massless, or Yukawa-like if the underlying field  transmitting this force between bodies has mass. Such a new force will produce a difference in the Sun's  acceleration of Earth and Moon, because the latter two bodies are of different compositions --- the Earth has a substantial iron core while the Moon is composed of low-Z mantle-like materials. The fractional difference in acceleration rates of Earth and Moon amounts to
\[
|\delta\vec{a}_{em}/\vec{g}_s|\;=\;\frac{K_s}{GM_S}\;\left(\frac{K_m}{M_m}-\frac{K_e}{M_e}\right)
\;(1+\mu R)\;e^{-\mu R}
\]
and it will supplement any difference of accelerations resulting from the possible anomalies in the bodies' gravitational to inertial mass ratios due to  gravitational self energies.  LLR has become a sufficiently precise tool for measuring $|\delta\vec{a}_{em}|$, it now competes favorably with ground-based laboratory measurements looking for composition-dependence of free fall rates; and LLR is the premier probe for measuring a body's $M(G)/M(I)$ ratio as given by Equation (2).

If Earth and Moon fall toward the Sun at different rates due to either of the mechanisms discussed, then the lunar orbit is polarized along the solar direction. Detailed calculation of this polarization reveals an interesting interactive feedback mechanism which acts between this $cos(D)$ polarization and the  $cos(2D)$ Newtonian solar tide perturbation of the lunar orbit, the {\it variation}). The result is an amplification of the synodic perturbation
\bea
\delta r(t)_{me}&=&\frac{3}{2}
\frac{\Omega}{\omega}\;R\;F(\Omega/\omega)\;\delta_{me}\;cosD \\
&\cong&2.9\;10^{12}\;\delta_{me}\;cosD\quad cm
\eea
with $\delta_{em}\;=\;|(\vec{a}_e-\vec{a}_m)/\vec{g}_s|$, $R$ is distance to the Sun, $\Omega$ and $\omega$ are the sidereal frequencies of solar and lunar motion, and $D$ is the lunar phase measured from new moon. The feedback amplification factor for the lunar orbit is already $F(\Omega/\omega)\cong1.75$; it grows further with larger orbits with an interesting resonance divergence for an orbit about twice the size as that of the Moon \cite{nor95, dmvk}.Computer integration of the complete Eq. (1) for the Sun-Earth-Moon system dynamics confirms these analytically estimated polarization sensitivities.

The most recent fits of the LLR data find no anomalies in the $cosD$ amplitude to precision of $4\;millimeters$, so from Equation (y) $\delta_{me}$ is constrained to be less than $1.3\;10^{-13}$. Neglecting any possible composition dependence, and using Equation () with an estimate for the fractional gravitational self energy of the Earth being $4.5\;10^{-10}$, a constraint on a combination of the two {\it Eddington} parameters
\be
|4\beta-3-\gamma|\;\leq\;4\;10^{-4}
\ee
If metric gravity is a combination of scalar and tensor interaction, the small number of this constraint is an approximate measure of the scalar interaction strength compared to the dominant tensor interaction.  One scenario which could explain today's weakness of the scalar interaction is illustrated in Figure R.  Scalar-tensor metric gravity involves one coupling function $V(\phi)$; the slope of this function gives the strength of the scalar interaction, and in combination with the function's curvature also determines gravity's $1/c^2$ order non-linearity; near an extremum of $V(\phi)$ the Eddington parameters are given by simple properties of the coupling function
\bea
1-\gamma\;&\cong&\;\frac{1}{2}\left(\frac{d\,lnV(\phi)}{d\phi}\right)^2 \\
\beta-1\;&\cong&\;\frac{1-\gamma}{8}\frac{d^2\,lnV(\phi)}{d\phi^2}
\eea
As the universe expands, the dynamical equations for the background scalar field will drive the scalar to a minimum of the coupling function, if it exists, and where $\gamma$ and $\beta$ take their general relativistic values.  Scalar gravity turns itself off naturally if an ``attractor'' exists in its coupling function $V(\phi)$.  But that process, being dynamical, should not be entirely complete today, and the small remnant of scalar interaction may still be detectable by sufficiently precise testing of relativistic gravity using LLR and other experiments \cite{dn1,dn2}.

\begin{figure}
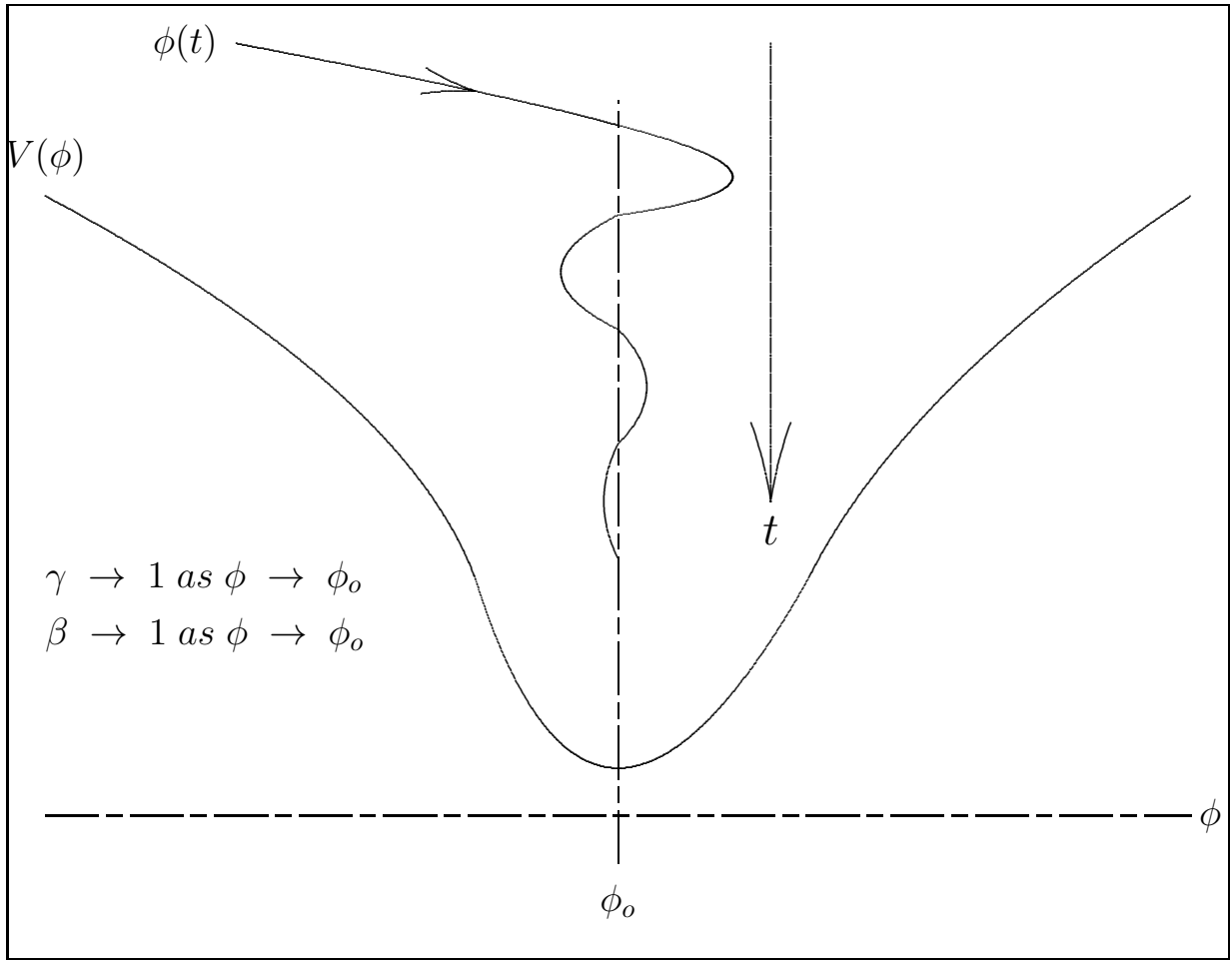

\beginpicture
\setcoordinatesystem units <1in,1in>
\setplotarea x from -3.2 to 3.2, y from -1 to 4
\putrectangle corners at -3.2 -1 and 3.2 4
\put {\large $\gamma\;\rightarrow\;1\;as\;\phi\;\rightarrow\;\phi_o$} [l] at -3 1
\put {\large $\beta\;\rightarrow\;1\;as\;\phi\;\rightarrow\;\phi_o$} [l] at -3 .7
\linethickness=10pt
\setquadratic \plot -3 3 -1.5 2 -.75 1 0 0 1 1 1.75 2 3 3 /
\linethickness=.4pt
\setquadratic \plot -2 3.8 -.75 3.55  0 3.37  .6 3.1  0 2.9  -.3 2.6  0 2.3  .15 2  0 1.7 -.075 1.4 0 1.1 /
\put {\large $V(\phi)$} at -3 3.2
\put {\large $\phi(t)$} [r] at -2.1 3.8
\put {\large $\phi_o$} at 0 -.7
\put {\large $\phi$} at 3.1 -.25
\arrow <30pt> [.2,.5] from .8 3.8 to .8 1.4
\arrow <20pt> [.2,.5] from -.8 3.56 to -.75 3.55
\put {\Large $t$} at .8 1.25
\setdashpattern <20pt, 3pt, 6pt, 3pt>
\putrule from 0 -.5 to 0 3.5
\putrule from -3 -.25 to 3 -.25
\endpicture
\caption{Typical cosmological dynamics of a background scalar field is shown if that field's coupling function $V(\phi)$ has an attracting point $\phi_o$. The strength of the scalar interaction's coupling to matter, proportional to the derivative of the coupling function, weakens as the attracting point is approached; so in a scalar-tensor metric theory, for example, the Eddington parameters $\gamma$ and $\beta$ both approach the pure tensor gravity values of one.} 
\end{figure}

The LLR result can also place limits on the spatial gradient of the {\it fine structure constant}, $\alpha=e^2/\hbar c$, in the proximity of the Sun.  If $\alpha$ is a function of a scalar field whose source includes ordinary matter, a spatial gradient of $\alpha$ near bodies should exist, and composition-dependent accelerations of other objects toward this body should occur
\[
\delta \vec{a}_i\;=\;-\,\frac{\partial lnM_i}{\partial ln\alpha}\;c^2\frac{\vec{\nabla}\alpha}{\alpha}
\]
The dominant electromagnetic contribution to the mass-energy of different elements is due to the electrostatic energy among the $Z$ nuclear protons.  This energy fractionally varies by an order of magnitude (from a few parts in $10^4$ to a few parts in $10^3$) as one proceeds through the periodic table from low-$Z$ to high-$Z$ elements.  For the Earth with its iron-core and Moon composed almost entirely of mantle-like materials, one can conclude from the LLR constraint on $\delta_{me}$ that any gradient of $\alpha$ due to and toward the Sun is quite small compared to the Sun's gravitational field $\vec{g}_s$
\[
\frac{c^2\,|\vec{\nabla}ln\alpha|}{|\vec{g}_s|}\;\leq\;4\,10^{-10}
\]
This should be compared with the best constraints on time variation of $\alpha$, which in units of the Hubble expansion rate are substantially weaker
\[
H\,\frac{\dot{\alpha}}{\alpha}\;\leq\;10^{-5}
\]
This suggests that unless there are unusual sources for the scalar field which controls the value of $\alpha$, e.g., sources which are present in an average cosmological context but which do not concentrate in ordinary matter, or other special situations, then today's LLR constraint on the spatial gradient of $\alpha$ is the significant present measure of the constancy of $\alpha$.  

\subsection*{Geodetic Precession of the Local Inertial Frame.}

Because Earth and Moon travel at different velocities through the Sun's gravitational field, terms from lines $D$ and $F$ of Equation () are present which accelerate the Moon relative to Earth.  A particularly interesting part of the relative acceleration is proportional to both $\vec{V}$ and $\vec{u}$ and the Sun's acceleration with, as shown in Figure 1, $\vec{V}$ being the velocity of the Earth relative to the Sun, and $\vec{u}$ the velocity of the Moon relative to Earth. These terms form deSitter's Coriollis-like acceleration 
\be
\delta \vec{a}_m = 2\;\vec{\Omega}_{dS}\times\vec{u}
\ee
with
\be
\vec{\Omega}_{dS} = \frac{2\gamma+1}{2}\;\frac{GM_s}{c^2R^3}\;
\vec{R}\times\vec{V}
\ee
and its geometrical interpretation is the local precession of the inertial frame at rate $\vec{\Omega}_{dS}$ which amounts to about $19.2\;mas/y$.  The effect of this perturbing acceleration on the orbit is primarily an additional rate of perigee precession with respect to distant inertial space. This is measured by comparing the Moon's anomalistic frequency $\dot{A}$ (rate of eccentric motion) with its synodic frequency $\dot{D}$ (rate of monthly phase), and with the latter converted into lunar sidereal frequency $\omega$ (orbital rate) by adding to $\dot{D}$ the annual rate $\Omega$ which is provided by results from other solar system experiments.  Sidereal minus anomalistic frequency of lunar motion includes deSitter's precessional rate as a supplement to the Newtonian tidal contributions to perigee precession. These lunar frequencies are measured from range signal perturbations whose size grows linearly in time. The Moon's range from Earth includes several dominant oscillatory contributions
\[
\delta r_{me}=L_{ecc}\;cos(A)\;+\;L_{var}\;cos(2D)\;+\;L_{evc}\;cos(2D-A)\;+\;...
\]
with $L_{ecc}$ being the amplitude of eccentric motion, $L_{var}$ being the amplitude of solar tidal perturbation called {\it variation}, and $L_{evc}$ being the amplitude of the hybrid {\it evection} perturbation due to both the solar tidal force and the eccentric motion of the Moon. The least-squares-fit of the LLR data, which yields best estimates for the two key lunar frequencies, will then involve the parameter 'partials'
\bea
\frac{\partial\delta r_{me}}{\partial\dot{A}}&=&
-t\;\left(L_{ecc}\;sin(A)\;-\;L_{evc}\;sin(2D-A)\right) \nonumber \\
\frac{\partial\delta r_{me}}{\partial\dot{D}}&=&
-2t\;\left(L_{var}\;sin(2D)\;+\;L_{evc}\;sin(2D-A)\right) \nonumber
\eea    
Measurement precision of the deSitter precession grows especially with total time of the LLR experiment, not only because of the growing quantity and quality of the accumulated range measurements, but also because of the linear growth in signal sensitivity. A most recent fit of the LLR data confirms presence of the geodetic precession with precision of $0.07\;mas/y$ \cite{wm99}.

\subsection*{Time Evolution of Gravity's Coupling Strength G.}

Time evolution of Newton's coupling parameter $G$ results in proportional evolutions for both the radial size and frequencies of the lunar motion. Slightly different orbital changes occur when a torque (indicated by $\dot{L}$) acts on the orbit 
\bea
\frac{\dot{r}}{r}\;&=&-\frac{\dot{G}}{G}+2
\frac{\dot{L}}{L} \nonumber \\
\frac{\dot{\omega}_n}{\omega_n}\;&=&\;2\frac{\dot{G}}{G}-3
\frac{\dot{L}}{L} \nonumber
\eea    
During the earlier years of the LLR experiment it has been the mean orbital radius signal
\[
\delta r(t)_{me}\;=\;\left(2\frac{\dot{L}}{L}-
\frac{\dot{G}}{G}\right)\;r\;(t-t_o)
\]
which has been used to measure $\dot{G}$.  But this involved estimating and subtracting a contribution to $\dot{r}$ which results from the orbital torque exerted on the Moon by the ocean tidal bulges on Earth which, because of friction, lag in angle from the direction toward the Moon. The inclination and 18.6 year precession of the lunar orbit's plane  result in a modulation of the tidal contribution to $\dot{r}$ which helps the separation of the two perturbations after accumulation of sufficient years of data. But the data set produced by LLR has in recent years become sufficiently extended in time so that the range signals associated with frequency shifts, which grow quadratically in time, are becoming dominant in the fit for $\dot{G}$.  Recall that the two lunar phases can be expanded in terms of initial phase, rate, and {\it acceleration} 
\bea
D(t)\;&=&\;D\;+\;\dot{D}\:(t-t_o)\;+\;\frac{1}{2}\ddot{D}\:(t-t_o)^2\;+\;...  \\ [.1in]
A(t)\;&=&\;A\;+\;\dot{A}\:(t-t_o)\;+\;\frac{1}{2}\ddot{A}\:(t-t_o)^2\;+\;... 
\eea
The synodic frequency is by definition equal to difference of lunar sidereal rate and Sun's sidereal rate around the Earth
\[
\dot{D}\;=\;\omega\:-\:\Omega
\]
while the Moon's {\it anomalistic} rate is derivable from the underlying equation of motion and can be expressed in the form
\[
 \dot{A}\;=\;\omega\;-\;\frac{3}{4}\frac{\Omega^2}{\omega}\;-\;\frac{225}{32}\frac{\Omega^3}{\omega^2}\;-\;...\;-(\gamma+1/2)\frac{GM}{c^2 R}\Omega\;+\;...
\]
which consists of the classical Newtonian expression plus relativistic modifications, with the dominant geodetic precession contribution shown.  From these two expressions the solar sidereal rate and its acceleration can then be expressed 
\bea
\Omega\;&=&\;\dot{A}-\dot{D}\;+\;\frac{3}{4}\frac{(\dot{A}-\dot{D})^2}{\dot{A}}\;+\;... \nonumber \\
\dot{\Omega}\;&=&\;\ddot{A}-\ddot{D}\;+\;... \nonumber
\eea
While the lunar phases $A$ and $D$ suffer accelerations due to any tidal torques acting between Earth and Moon, the solar rate $\Omega$ is not affected by the tidal torques.  Acceleration of this solar rate is therefore a rather pure measure of a time variation of $G$.  Noting from Equations (15-16) that the partials for $\ddot{A}$ and $\ddot{D}$ will grow in amplitude quadratic in time
\bea
\frac{\partial R_{calc}}{\partial \ddot{A}}\;&=&\;\frac{1}{2}t^2\,\left(L_{ecc}\,cos(A)\,-\,L_{evc}\,cos(2D-A)\right) \\
\frac{\partial R_{calc}}{\partial \ddot{D}}\;&=&\;t^2\,\left(L_{var}\,cos(2D)\,+\,L_{evc}\,cos(2D-A)\right)
\eea
it follows that the formal error in measuring $\dot{G}$ decreases as the inverse square of the time span $T$ of LLR observations.  For a uniform time distribution of observations one obtains
\[
with\hspace{.3in}\frac{\dot{G}}{G}\;=\;\frac{1}{2}\frac{\dot{\Omega}}{\Omega}\hspace{.4in}\left(\frac{\delta\dot{G}}{G}\right)_{RMS}\;=\;\sqrt{\frac{360}{N}}\:\frac{1}{\Omega\:T^2}\:\frac{\sigma}{\sqrt{4L_{var}^2+3L_{evc}^2}}
\]
with $\sigma$ being the rms size of individual range measurement errors, and $N$ is the total number of measurements spread over the time $T$. A recent fit of almost 30 years of  LLR data yields the excellent measurement constraint \cite{wm99}
\be
\frac{\dot{G}}{G}\;\cong\;(0\;\pm\;1.1)\;10^{-12}\;y^{-1}\hspace{.4in} 
\ee
This amounts to about 1/60 the observed Hubble expansion rate of the universe. And with this measurement precision now growing quadratic in time, LLR should continue indefinately being at the cutting edge in supplying the measurement of $\dot{G}$.

\section{Additional Yukawa Interaction?}

When the supplementary interaction given by Eq. (7) is of a Yukawa nature, $\mu\neq 0$, it contributes to precession of  periastron for a near-circular orbit of radius $r$ in amount
\[
\frac{\delta(\omega-\omega_o)}{\omega}\;=\;\frac{1}{2}\frac{K_iK_j}{GM_iM_j}\;(\mu r)^2\;exp(-\mu r)
\]
with $\omega$ and  $\omega_o$ being the orbit's sidereal and eccentric frequencies, respectively. And this perturbation of precession rate also occurs if the Yukawa force is metric, $K_i\sim M_i$, or non-metric. With the Moon's perigee precession rate measured to precision $.07\;mas/y$ and showing no anomaly, then for Yukawa ranges in the vicinity of that for maximum sensitivity of lunar perturbation, $\mu r=2$, the strength of the Yukawa force is decisively constrained
\[
\frac{|K_eK_m|}{GM_eM_m}\;\leq\;5\;10^{-12}\quad\left(\frac{4}{(\mu r)^2}\;exp(\mu r-2)\right)
\]

\section{Gravitomagnetism.}

Line $C$ of the complete N-body gravitational equation of motion given by Eq. (1) indicates a post-Newtonian gravitational force proportional to the velocities of both bodies in the interaction, and in analogy with electromagnetic theory, it has been called the {\it gravitomagnetic} interaction.  From line $C$ of Eq. (1), this acceleration is
\[
\delta\vec{a}_i\;=\;(2+2\gamma)\;\sum_{j\neq i}
\frac{Gm_j}{c^2r_{ij}^3}\left(\vec{r}_{ij}\,\vec{v}_i\cdot\vec{v}_j\,-\,\vec{r}_{ij}\cdot\vec{v}_i\,\vec{v}_j\right)
\]
It often has been claimed that the presence of gravitomagnetism within the total gravitational interaction has not been experimentally confirmed and measured. Indeed, different experiments have been under development to explicitly observe the effects of this historically interesting prediction of general relativity. But this gravitomagnetic acceleration already plays a large role in producing the final shape of the lunar orbit, albeit in conjunction with the rest of the total equation of motion; the precision fit of the LLR data indicates that gravitomagnetism's presence and specific strength in the equation of motion can hardly be in doubt. Because both the Earth and Moon are moving in the solar system barycentric frame --- the frame in which the dynamical equations are formulated and then integrated into orbits --- a gravitomagnetic interaction exists between these two bodies, the Earth having velocity $\vec{V}(t)$ and the Moon's being $\vec{V}(t)+\vec{u}(t)$,  as seen in Figure (2).  As a result of these mutual motions, perturbations to the Earth-Moon range from the gravitomagnetic acceleration result proportional to both $V^2$ and $Vu$
\bea
\delta r(t)\;&\cong&\;\frac{Gm_e}{r^2}\left(-\frac{4}{3\omega^2}\frac{V^2}{c^2}\quad cos(2D) \;+\;
\frac{2}{\omega\Omega}\frac{Vu}{c^2}F(\Omega/\omega) \quad cos(D)\right) \nonumber \\
&\cong&\;-\; 530\;cos(2D)\; +\;525\; cos(D)\quad cm
\eea       
As previously discussed, the amplitudes of the lunar motion at both these periods (monthly and semi-monthly) are determined to better than half a centimeter precision in the total orbital fit to the LLR data. It would be impossible to understand this fit of the LLR data without the participation of the gravitomagnetic interaction in the underlying model, and with strength very close to that provided by general relativity, $\gamma=1$. As in electromagnetic theory, the velocity-dependent force terms on lines $C$ and $D$ of Eq. (6) can individually be changed by formulating the dynamics in different frames of reference, but the very ability to reformulate the equations of motion in different frames without introducing new frame-dependent terms depends on the local Lorentz invariance (LLI) of gravity. It is the entire package of velocity-dependent, post-Newtonian terms which includes the gravitomagnetic terms, lines $C$ plus $D$ of Eq. (6), that produces the LLI; the {\it Eddington} parameter $\gamma$ represents the only freedom in the structure of this LLI package. Our confidence in the exhibited structure of this total collection of velocity-dependent terms is established in proportion to the precision with which the various {\it preferred frame}, LLI-violating effects in the solar system proportional to $W^2$, $WV$, and $Wu$ have been found to be absent \cite{norw}.  LLR has been one of the main contributors in establishing gravity's LLI through null measurements of several $W$-dependent effects \cite{nw72,nor73,mnv,nor96}.       

\section{Inductive Inertial Forces.}

{\it Inductive} forces are shown on line $E$ of Eq. (1); in such forces the acceleration of one mass element induces an acceleration of another proximite mass element (e.g., $i$ and $j$ in Figure 2). From line $E$ of Eq. (1) we have
\be
\delta\vec{a}_i\;=\;
\sum_{j\neq i}\frac{Gm_j}{2c^2r_{ij}}\left((4\gamma +3)\vec{a}_j\;+\;
\vec{a}_j\cdot\hat{r}_{ij}\hat{r}_{ij}\right)
\ee
These accelerations play a key part in altering the inertial masses of the Earth and Moon because of their internal gravitational binding energies; either the absence or an anomalous strength of these inductive forces would translate directly into differences between the acceleration rates of these whole bodies toward the Sun.  A polarization of the Moon's orbit in the solar direction, as previously discussed, would result. The forces, Eq. (21), acting between the mass elements of Earth, for example, by themselves would lead to an anomalous polarization of the lunar orbit of very large magnitude
\be
\delta r(t);\cong\;130\quad\cos(D)\quad meters
\ee
Only when these inductive forces are combined with the other post-Newtonian inertial forces shown on line $F$ of Eq. (1) does the total inertial self force of a body become
\bea
\delta\vec{f}\;=\;&-&\:\frac{1}{c^2}\left(\frac{1}{2}\sum_i m_i v_i^2\;-\;
\frac{G}{2}\sum_{i,j}\frac{m_i m_j}{r_ij}\right)\;\vec{a} \nonumber \\
&-&\;\frac{1}{c^2}\left[\sum_i m_i \vec{v}_i\;\vec{v}_i\;-\;\frac{G}{2}
\sum_{i,j}\frac{m_i m_j}{r_{ij}^3}\vec{r}_{ij}\:\vec{r}_{ij}\right]\:\cdot\:\vec{a} \nonumber
\eea
The first line of this total self force is now the expected inertial force due to the internal kinetic energy and gravitational binding energy within the body. The second line represents contributions to the body's internal {\it virial} which, when totaled over all internal force fields, vanishes for a body in internal equilibrium and experiencing negligible external tidal-like forces.  These  self forces of a body are an integral part of the determination of the total gravitational to inertial mass ratio of bodies discussed previously, and in general relativity they are cancelled by equal contributions of internal energies to a body's gravitational mass. They were explicitly discussed here in order to show the large size of such inductive force contributions which must necessarily be taken into account in the fit of theory to the LLR data.

{\bf This work has been supported by the National Aeronautics and Space Administration through contract NASW-97008 and NASW-98006.}

\end{document}

%% file: prepicte.tex
\catcode`@=11 \catcode`!=11

\expandafter\ifx\csname fiverm\endcsname\relax
  \let\fiverm\fivrm
\fi
  
\let\!latexendpicture=\endpicture 
\let\!latexframe=\frame
\let\!latexlinethickness=\linethickness
\let\!latexmultiput=\multiput
\let\!latexput=\put
 
\def\@picture(#1,#2)(#3,#4){%
  \@picht #2\unitlength
  \setbox\@picbox\hbox to #1\unitlength\bgroup 
  \let\endpicture=\!latexendpicture
  \let\frame=\!latexframe
  \let\linethickness=\!latexlinethickness
  \let\multiput=\!latexmultiput
  \let\put=\!latexput
  \hskip -#3\unitlength \lower #4\unitlength \hbox\bgroup}

\catcode`@=12 \catcode`!=12

%% file: postpict.tex
\catcode`@=11 \catcode`!=11
  
\let\!pictexendpicture=\endpicture 
\let\!pictexframe=\frame
\let\!pictexlinethickness=\linethickness
\let\!pictexmultiput=\multiput
\let\!pictexput=\put

\def\beginpicture{%
  \setbox\!picbox=\hbox\bgroup%
  \let\endpicture=\!pictexendpicture
  \let\frame=\!pictexframe
  \let\linethickness=\!pictexlinethickness
  \let\multiput=\!pictexmultiput
  \let\put=\!pictexput
  \let\input=\@@input   
  \!xleft=\maxdimen  
  \!xright=-\maxdimen
  \!ybot=\maxdimen
  \!ytop=-\maxdimen}

\let\frame=\!latexframe

\let\pictexframe=\!pictexframe

\let\linethickness=\!latexlinethickness
\let\pictexlinethickness=\!pictexlinethickness

\let\\=\@normalcr
\catcode`@=12 \catcode`!=12